\documentclass[aps,pra,preprint,superscriptaddress]{revtex4-1}
\usepackage{graphicx} 
\usepackage{amsmath}
\usepackage{bbold}

\begin{document}

\title{A naturally stable Sagnac-Michelson nonlinear interferometer}

\author{Joseph M. Lukens}\email{lukensjm@ornl.gov}\affiliation{Quantum Information Science Group, Computational Sciences and Engineering Division, Oak Ridge National Laboratory, Oak Ridge, Tennessee 37831, USA}
\author{Nicholas A. Peters}\affiliation{Quantum Information Science Group, Computational Sciences and Engineering Division, Oak Ridge National Laboratory, Oak Ridge, Tennessee 37831, USA}\affiliation{The Bredesen Center for Interdisciplinary Research and Graduate Education, The University of Tennessee, Knoxville,  Tennessee 37996, USA }
\author{Raphael C. Pooser}\affiliation{Quantum Information Science Group, Computational Sciences and Engineering Division, Oak Ridge National Laboratory, Oak Ridge, Tennessee 37831, USA}\affiliation{Department of Physics, The University of Tennessee, Knoxville,  Tennessee 37996, USA}

\date{November 1, 2016}



\begin{abstract}
Interferometers measure a wide variety of dynamic processes by converting a phase change into an intensity change.  Nonlinear interferometers, making use of nonlinear media in lieu of beamsplitters, promise substantial improvement in the quest to reach the ultimate sensitivity limits.  Here we demonstrate a new  nonlinear interferometer utilizing a single parametric amplifier for mode mixing---conceptually, a nonlinear version of the conventional Michelson interferometer with its arms collapsed together. We observe up to 99.9\% interference visibility and find evidence for noise reduction based on phase-sensitive gain. Our configuration utilizes fewer components than previous demonstrations and requires no active stabilization, offering new capabilities for practical nonlinear interferometric-based sensors.
\end{abstract}


\maketitle

Interferometers have been and remain among the most useful and sensitive scientific tools.  From dispelling the theory of electromagnetic ether~\cite{michelson1887relative} to confirming the existence of gravitational waves~\cite{abbott2016observation}, they have proven crucial for advancing our understanding of the universe.  As such, their limitations and improvement remain a subject of much study.  The most common interferometers split input light into two modes, one of which contains a phase-changing system to be probed.  After interaction with the system, the two modes are recombined on a beamsplitter and the output modes detected. Fundamentally, the sensitivity of such an interferometer to detect a phase change is related not only to the number of photons used ($N$), but also to their quantum state~\cite{Caves1981}.  For coherent states of light, such as those produced by stable lasers, the phase uncertainty scales as the standard quantum limit, $1/\sqrt{N}$---a consequence of the shot noise for random particles \cite{Caves1981,jaekel1990quantum}. However, it has been shown that the standard quantum limit can be exceeded by injecting squeezed light, with the potential to reach the Heisenberg limit, where the uncertainty scales as $1/N$~\cite{Caves1981,jaekel1990quantum}.  

Yet reducing the noise through squeezing is not the only route toward the Heisenberg limit. One can also increase the signal relative to the noise with a nonlinear interferometer (NLI)---alternately called the SU(1,1) interferometer due to the interaction's representation in the group SU(1,1)~\cite{yurke1986,Plick2010,ou2012enhancement}. In an NLI, the linear beamsplitters are replaced by nonlinear optical parametric amplifiers. Even with coherent state inputs, significant sensitivity boosts are possible at any power level~\cite{Plick2010}. In our case, we seed one port of the nonlinear beamsplitter with a coherent state (the probe), amplifying it independent of input phase, and stimulating a conjugate mode in the process. In the second amplification, the gain becomes phase-sensitive and, in principle, adds the minimum noise allowed by quantum mechanics~\cite{caves1982}.  As the phase-sensitive gain can be high relative to the added noise, there is the potential to improve the signal-to-noise ratio (SNR) via quantum noise cancellation~\cite{xin2016}. Thus, compared to a linear interferometer,  interference fringe maxima are increased even though the same number of photons are used to probe the phase change.  In the high-gain limit, this leads to an SNR improvement of approximately twice the intensity gain~\cite{ou2012enhancement}.  Further, it has been shown theoretically that NLIs tolerate detection loss better than linear interferometers utilizing squeezed inputs~\cite{ou2012enhancement,marino2012effect}.  To date, there have been several experimental demonstrations of parametric-amplifier-based NLIs~\cite{xin2016, jing2011realization,kong2013NLIvis,hudelist2014quantum,wang2015experimental}, as well as explicit experimental characterization of the underlying nonlinear beamsplitter~\cite{Fang:16}.  

\begin{figure*}[t!]
\centering
\includegraphics[width=6.5in]{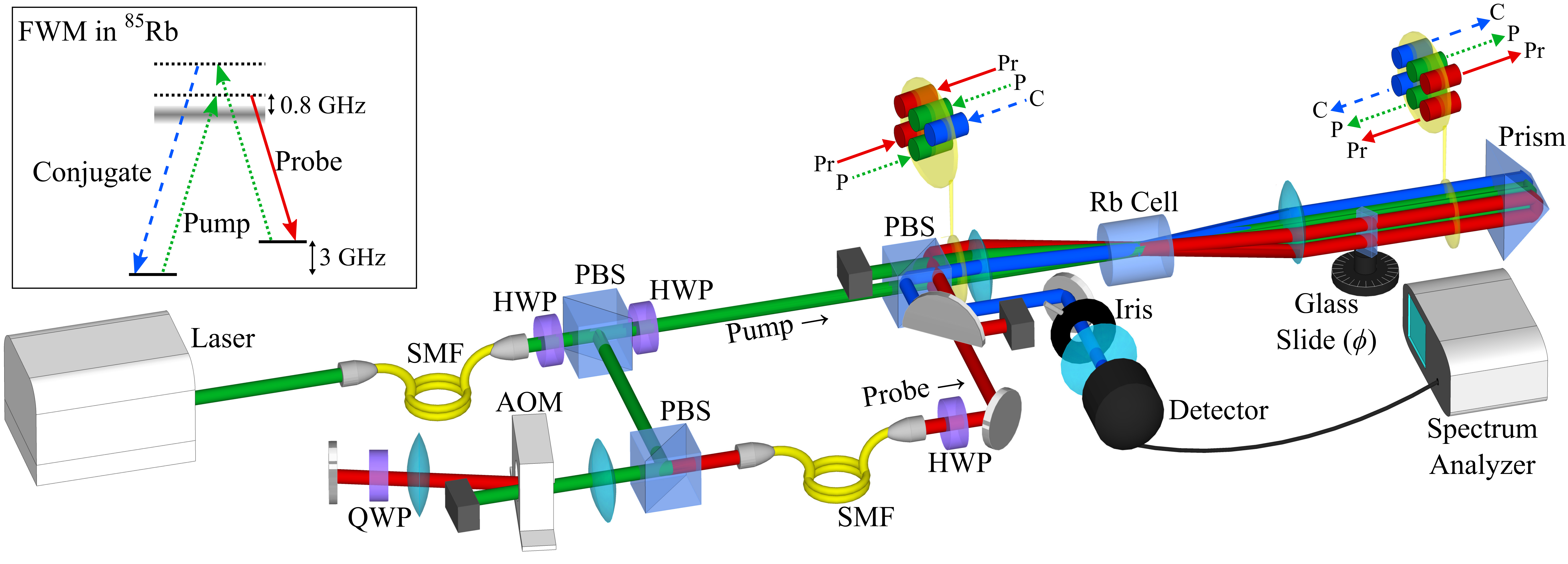}
\caption{Experimental setup illustrating the general concept of Sagnac-Michelson NLI. AOM: acousto-optic modulator; HWP: half-wave plate; PBS: polarizing beamsplitter; QWP: quarter-wave plate; SMF: single-mode fiber. Probe (Pr) and pump (P) fields produce a conjugate (C) beam in the parametric amplifier (Rb cell). On the return pass, the gain varies with the phase shift $\phi$ imparted by the glass slide, so that detecting either probe or conjugate output gives a measure of the induced phase shift. Zoomed-in slices mark the positions of the forward- and back-propagating fields, both outside and within the interferometer.}
\label{fig1}
\end{figure*}

However, these NLIs have relied upon a Mach-Zehnder configuration, and must be actively stabilized for practical use as a sensor~\cite{wang2015experimental}.  Here we report the realization of a parametric-amplifier-based nonlinear interferometer which instead utilizes a single nonlinear element traversed twice, like the beamsplitter in a linear Michelson. Consequently, the sensing field passes through the phase-shifting element twice and accumulates double the phase compared to previous NLI approaches, further improving its usefulness as a metrological tool. But in contrast to a conventional Michelson, the beams travel along near-parallel paths, yielding intrinsic stability~\cite{Note}.  With the exception of the phase-shifting element, all fields make contact with the same components, similar to a displaced-Sagnac interferometer as in, e.g.,~\cite{Kwiat1999,Hosten2006,Nagata2007}. We use the term ``Sagnac- Michelson'' to describe this physical realization uniting the advantages of both interferometer types. We note that these concepts (single nonlinear medium, common-path stability) have been applied in experiments with Raman scattering as the nonlinear interaction~\cite{Chen2015}, but not in the context of parametric amplification and NLIs.

In this Letter, we report operation of this new, ultrastable Sagnac-Michelson NLI. We also demonstrate for the first time the advantages of spatial mode filtering---common in linear interferometry---to boost fringe contrast in a nonlinear interferometer. We attain visibilities up to $(99.93\pm 0.01)\%$ with modest spatial filtering and modulation of the measured beam so that it may be detected far away from low-frequency background noise; both techniques will likely prove crucial in realizing practical optical sensors approaching the Heisenberg limit.

The experimental setup for our stable NLI is outlined in Fig.~\ref{fig1}. In a linear Michelson, collapsing the arms to improve stability is not usually an option due to the angles of incidence: free-space beamsplitters generally do not operate well at oblique angles.  However, the nonlinear Michelson is particularly well suited to merging the arms, since the nonlinear beamsplitter can be configured with beams that propagate closely in free space (see Fig.~\ref{fig1}). A powerful pump field and relatively weak probe seed are focused into a parametric amplifier (here, Rb cell), generating a conjugate field in the process.  Next, one of the three fields (in this case, the probe) experiences a phase shift before all three reflect back into the amplifier for phase-sensitive mixing. Finally, the conjugate field is detected, and its intensity is related to the phase shift $\phi$. While alternative detection schemes are possible, such as summing the probe and conjugate outputs~\cite{Plick2010} or homodyne detection with a local oscillator~\cite{ou2012enhancement}, the general behavior is the same.

In our experiment, the nonlinear process is four-wave mixing (FWM) in a hot $^{85}$Rb vapor~\cite{mccormick2007, McCormick2008}. Yet the approach is entirely generic; for example, using widely wavelength-nondegenerate FWM, a waveguided nonlinear interferometer can be configured using wavelength-division multiplexing such that all optical signals share the same spatial mode.  Thus we expect the nonlinear Sagnac-Michelson configuration to find many applications in other platforms which support phase-sensitive parametric amplification~\cite{Tong:13}, for example in fiber-optics~\cite{Agarwal:14, Dailey:15} and on-chip photonic waveguides~\cite{Neo:13,Zhang:14}.  

In the FWM configuration used here, a narrow-linewidth Ti:Sapphire laser at $\sim$795 nm is coupled into optical fiber for mode clean up and split into two arms on a polarizing beamsplitter (PBS), with the split ratio controlled by a preceding half-wave plate (HWP). The more powerful horizontal ($H$) mode serves as the pump, and the vertically ($V$) polarized output is sent through an acousto-optic modulator (AOM) in a double-pass configuration~\cite{Donley2005}. The microwave drive signal at 1.52 GHz carrier redshifts the beam by 3.04 GHz in the double pass.  An additional 750 kHz amplitude modulation acts as a chopper on the probe, shifting the measured signal away from electronic noise at low frequencies. The returning redshifted probe is again spatially filtered by an optical fiber before entering the NLI.

The pump and the probe are combined on another PBS and then focused by a 50 cm focal-length lens with a crossing angle of 8~mrad into a 12.7 mm long $^{85}$Rb vapor cell heated to $\sim$120$^\circ$ C. The pump (and consequently probe) wavelength is scanned and locked to maximize FWM gain; see the inset in Fig.~\ref{fig1} for an energy-level diagram. This process has been used in previous Mach-Zehnder NLI demonstrations~\cite{jing2011realization,kong2013NLIvis,hudelist2014quantum,wang2015experimental}, as well as in applications including generation of squeezing~\cite{mccormick2007,Qin2014}, quantum imaging~\cite{boyer2008}, surface plasmon resonance sensors~\cite{Lawrie2013, pooser2015plasmonic}, compressive quantum imaging~\cite{Lawrie2013_2}, and sub-shot-noise displacement measurements~\cite{Pooser:15}.  

\begin{figure}[b!]
\includegraphics[width=3.4in]{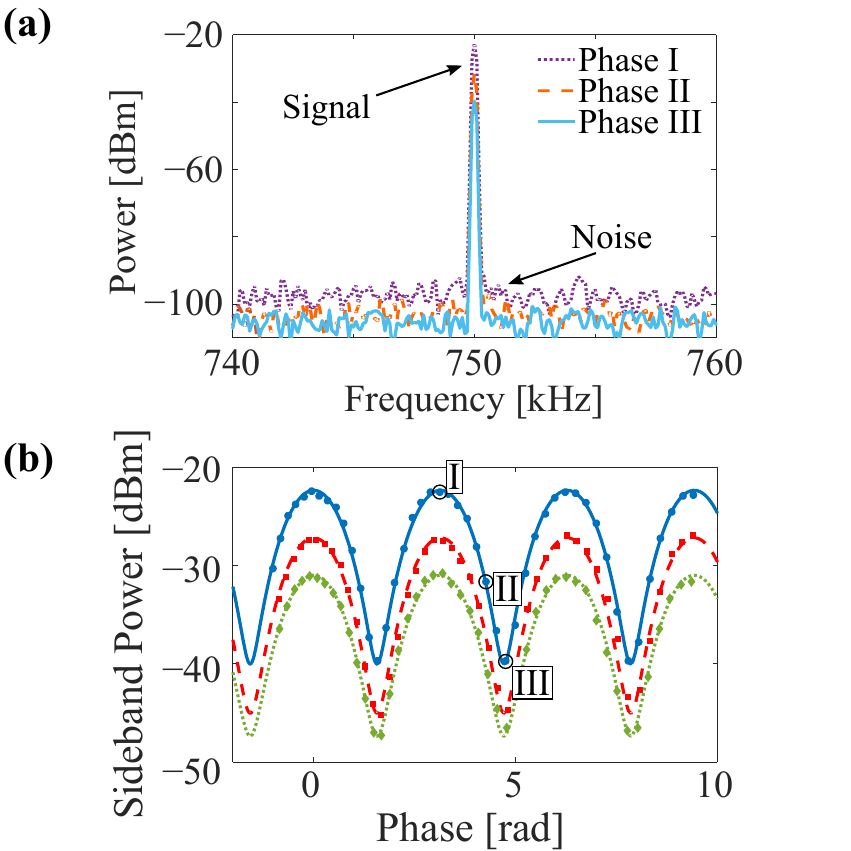}
\caption{Experimental results without spatial filtering. (a) Example microwave spectra for three phase settings, showing the signal as a peak at 750 kHz and the noise as the neighboring background. (b) Interference fringe measurements. Peak sideband power as a function of phase shift, for forward single-pass gains of 4.0 (circles/solid blue), 3.4 (squares/dashed red), and 2.8 (diamonds/dotted green). The Roman numerals relate the measurements in (a) to the resulting locations in (b).}
\label{fig2}
\end{figure}

The pump, probe, and newly generated conjugate are reimaged into the cell in a 4$f$-system, with the probe phase-shifted by a rotatable glass slide. Additionally, all three beams are vertically displaced with a right-angle prism retroreflector, akin to the horizontal translation typically associated with linear displaced-Sagnac interferometers. Such displacement ensures that the return fields propagate in different spatial modes through the $^{85}$Rb vapor, suppressing unwanted nonlinear interactions between forward- and back-propagating fields and enabling easy separation of the output fields from the inputs. (See the two cross-sections in Fig. \ref{fig1} for detailed beam positions.)  The $V$-polarized probe and conjugate outputs are then picked off by a D-shaped mirror; the conjugate is detected, and its photocurrent is fed into a microwave spectrum analyzer. Figure \ref{fig2}(a) plots examples of the measured tone at 750 kHz for three different phase settings. In this way, the measured sideband lies in a low-noise region of the microwave spectrum and within the quantum noise reduction bandwidth, a significant improvement over previous experiments in which the measured signal was a DC voltage~\cite{jing2011realization,kong2013NLIvis,hudelist2014quantum}. For example, in order to utilize the quantum enhancement observed at 1 MHz offset in~\cite{hudelist2014quantum}, the phase-sensing signal must be shifted away from DC, which our modulation scheme accomplishes.

Since we measure microwave power, the signal peak satisfies $P_\mathrm{sideband} = K\langle\hat{n}_c^2\rangle$, where $\hat{n}_c$ is the conjugate photon-number operator and $K$ is a constant related to detector response, amplifier gain, and spectrum analyzer settings. The noise is proportional to the variance: $P_\mathrm{noise}= K\langle\Delta\hat{n}_c^2\rangle$. Assuming zero phase mismatch, undepleted pump, equal intensity gains for each pass ($G_1=G_2=G$), and an input probe coherent state with mean photon number $|\alpha|^2\gg 1$, analysis similar to \cite{marino2012effect,ou2012enhancement} yields
\begin{equation}
\label{e1}
\langle\hat{n}_c^2\rangle=4G(G-1)|\alpha|^2 \cos^2\phi \left[ 1 + 4G(G-1)|\alpha|^2 \cos^2\phi\right]
\end{equation}
and
\begin{equation}
\label{e2}
\begin{split}
\langle\Delta\hat{n}_c^2\rangle & = 4G(G-1)|\alpha|^2 \cos^2\phi \left[ 1 + 8G(G-1) \cos^2\phi\right] \\
 & = \langle\hat{n}_c\rangle \left[1 + \frac{2}{|\alpha|^2}\langle\hat{n}_c\rangle \right].
\end{split}
\end{equation}
Here $\phi$ is the single-pass phase shift induced by the device under test, up to an overall constant. From Eq. (\ref{e2}), the statistics of the output conjugate field are super-Poissonian ($\langle\Delta\hat{n}_c^2\rangle > \langle\hat{n}_c\rangle$). In the limit where $\langle\hat{n}_c\rangle \rightarrow 0$ (the dark fringe), the statistics approach Poissonian.  Here the NLI yields its maximal SNR enhancement: a factor of $2G$ over a linear interferometer with the same phase-sensing photon number \cite{ou2012enhancement}. For unequal gains, the aforementioned expressions are more complicated, though the functional dependencies remain the same; accordingly, for fitting our experimental results we employ the form $P_\mathrm{sideband}^\mathrm{(FIT)} = C_0 + C_1 \cos^2\phi + C_2 \cos^4\phi$, to extract the non-unity visibility.

NLI interference fringe measurements without spatial filtering are shown in Fig. \ref{fig2}.   For each scan, the glass slide is incrementally rotated by a motorized stage, and the sideband power spectrum is measured at every phase setting.  Figure \ref{fig2}(a) shows the measured spectra at three example phase settings, which result in the points corresponding to the respective Roman numerals in Fig. \ref{fig2}(b). The curves show best fits based on $P_\mathrm{sideband}^\mathrm{(FIT)}$. Each set of data was acquired at different input pump powers, corresponding to first-pass intensity gains ($G_1$ values) of 4.0 (circles/solid blue), 3.4 (squares/dashed red), and 2.8 (diamonds/dotted green). The fitted visibilities for all three cases are 95--97\% within error bars. The duration of a single phase scan is on the order of 30 s. For such a large free-space interferometer (200 cm round trip), the setup is remarkably stable over these relatively long timescales. These results did not require a box around the interferometer, active stabilization, or thermal isolation.

\begin{figure}[t!]
\includegraphics[width=3.4in]{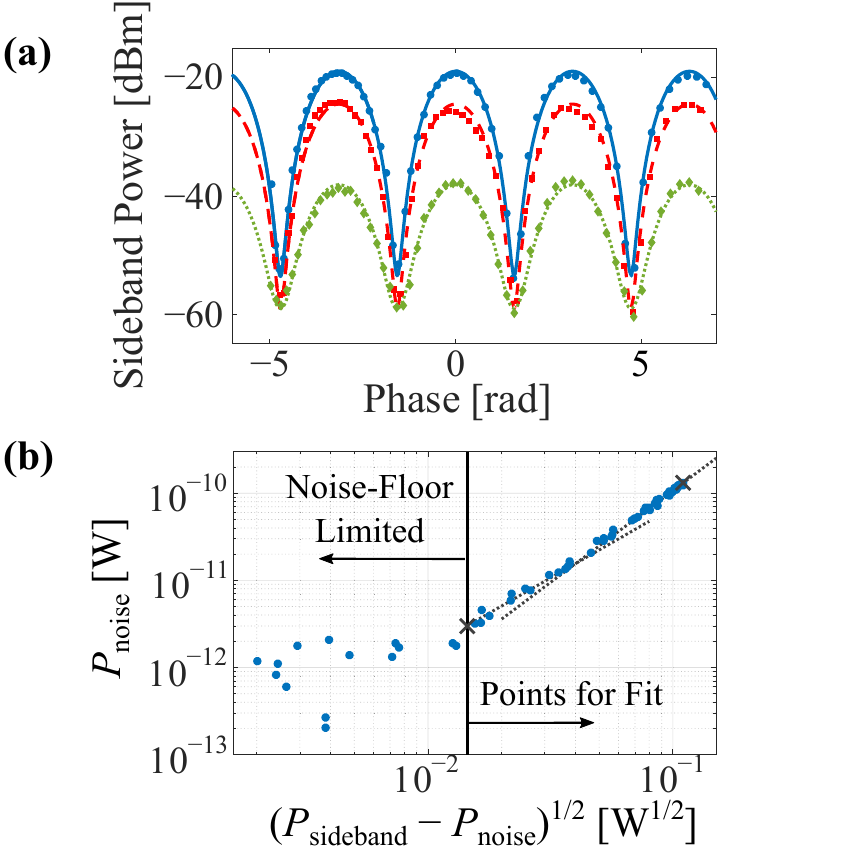}
\caption{Experimental results with spatial filtering. (a) Interference fringe measurements. Sideband power as a function of phase for three single-pass gain settings: 4.1 (circles/solid blue), 3.5 (squares/dashed red), and 2.5 (diamonds/dotted green). (b) Noise scaling of the 4.1 gain results from (a). The optical noise power is plotted against the root of the signal contrast, a quantity proportional to photon number; the dotted lines give the numerically fitted tangents at the two extrema.}
\label{fig3}
\end{figure}

In order to further boost interference visibility, we next consider spatial filtering of the output beam with an adjustable iris.  Figure \ref{fig3}(a) shows the sideband power fringes for three different gain values when the iris filters down to $\sim$6\% beam transmission. As before, each curve corresponds to different first-pass gains: 4.1  (circles/solid blue), 3.5 (squares/dashed red), and 2.5 (diamonds/dotted green). The interference contrast is significantly enhanced, with the solid curve (furthest from the noise level) yielding a visibility of $(99.93\pm0.01)\%$. The improvement is most likely due to the fact that the phase-sensing field is amplified across multiple spatial modes during the FWM process. The multispatial-mode nature, along with imperfections in beam alignment, may introduce nonuniform phase fronts during the amplification process. It has also recently been shown that amplitude modulation may introduce phase noise during phase-sensitive amplification~\cite{li2016}. These effects combined with the increased visibility obtained when applying a spatial filter suggest that structuring the pump and probe fields (such as using a top-hat pump mode) and tuning AOM alignment (the source of our amplitude modulation) would lead to increased visibility while simultaneously relaxing the need for spatial filtering on the output of the NLI.

Finally, as a primary goal is to improve the SNR, we study the noise behavior in our Sagnac-Michelson NLI. According to Eq.~(\ref{e2}), the theoretical output of our interferometer is characterized by a photon-number variance which scales linearly with the mean at a dark fringe and quadratically at a bright fringe. Since we employ direct detection of the conjugate mode, we expect technical noise to prevent observation of ideal performance as dark fringe power approaches zero. Nonetheless, a reduction in noise scaling near the dark fringes offers evidence of the quantum noise cancellation ultimately required for enhanced sensing. To this end, we examine in more detail the set of spectrum-analyzer traces corresponding to the circles/solid-blue scan in Fig.~\ref{fig3}(a). At each phase we calculate the noise level by averaging the spectral power adjacent to the 750 kHz sideband and subtract off the electronics floor to give the quantity $P_\mathrm{noise}$. Taking the root of the difference between the sideband peak and this noise level at each step produces an expression proportional to the mean photon number, since by Eqs. (\ref{e1}) and (\ref{e2}), $\sqrt{P_\mathrm{sideband}-P_\mathrm{noise}} = \sqrt{K} \langle\hat{n}_c\rangle$. Thus in a graph of $P_\mathrm{noise}$ against $\sqrt{P_\mathrm{sideband}-P_\mathrm{noise}}$, a log-log slope of 1 indicates Poissonian statistics, whereas a slope of 2 corresponds to the super-Possoinian scaling expected at the fringe maxima.

Figure \ref{fig3}(b) plots these two quantities against each other. Because the noise level reaches the electronics floor at low powers (left of the vertical solid line), we neglect these in the numerical fit. Plotted are the tangents to the polynomial fit at the upper and lower extrema of the the fit domain. The log-log slope increases from ($1.6\pm0.3$) to ($2.1\pm0.2$) as we move from dark to bright fringes, a signature of noise cancellation near the dark fringe. The scaling can be more directly probed at the dark fringe using more complex measurement techniques such as homodyne detection. We plan to implement other detection schemes, along with ensuring minimal excess noise on the input state, in future studies of this new interferometer.

In summary, we have realized the first inherently stable nonlinear interferometer, utilizing only a single parametric amplifier as beamsplitter and inducing a double phase shift in the sensing path. Not only does the single $^{85}$Rb cell reduce cost, size, weight and power requirements; it also automatically ensures matched nonlinear vapor cell properties in both passes. Measured visibilities in excess of 99.9\% demonstrate exceptional performance, and we uncover evidence of the expected noise reduction.  This proof-of-principle experiment demonstrates a new class of ultrastable nonlinear interferometer, which in the future could be implemented in single-mode fiber or other waveguide platforms for realizing practical and compact nonlinear-enhanced sensors.

\bigskip
\textsf{\textbf{Funding.}} Oak Ridge National Laboratory (Laboratory Directed Research and Development Program).
\bigskip

\textsf{\textbf{Acknowledgement.}} We thank E. M. Layden for technical assistance. This manuscript has been authored by UT-Battelle, LLC under Contract No. DE-AC05-00OR22725 with the U.S. Department of Energy.


\begin{thebibliography}{10}
\bibitem{Note}
If for practical reasons, additional optics are needed to reach a phase shifter far from the interferometer paths, the stability will be reduced.
\end{thebibliography}
\end{document}